\documentclass[reviewcopy]{elsart}

\usepackage{graphicx}
\usepackage{amsfonts,amsmath}

\newcommand{\ket}[1]{|#1\rangle}
\newcommand{\bra}[1]{\langle #1|}
\begin{document}
\begin{frontmatter}
\title{Nodal free geometric phases: Concept and application to 
geometric quantum computation}
\author[x]{Marie Ericsson}, 
\ead{M.Ericsson@damtp.cam.ac.uk} 
\author[y]{David Kult}, 
\ead{david.kult@kvac.uu.se}
\author[y]{Erik Sj\"oqvist\corauthref{fi}}, 
\corauth[fi]{Corresponding author.}
\ead{eriks@kvac.uu.se}
\author[x]{Johan {\AA}berg}
\ead{J.Aberg@damtp.cam.ac.uk}
\address[x]{Centre for
Quantum Computation, Department of Applied Mathematics and Theoretical
Physics, University of Cambridge, Wilberforce Road, Cambridge CB3 0WA,
United Kingdom}
\address[y]{Department of Quantum Chemistry, Uppsala University, 
Box 518, Se-751 20 Uppsala, Sweden}
\date{\today}
\begin{abstract}
Nodal free geometric phases are the eigenvalues of the final 
member of a parallel transporting family of unitary operators. 
These phases are gauge invariant, always well-defined, and can 
be measured interferometrically. Nodal free geometric phases 
can be used to construct various types of quantum phase gates. 
\end{abstract}
\begin{keyword}
Geometric phase; Quantum gates; Interferometry  
\PACS 03.65.Vf, 03.67.Lx, 07.60.Ly   
\end{keyword}
\end{frontmatter} 
\maketitle 
Geometric phases in noncyclic evolution \cite{samuel88} are undefined
if the initial and final states are orthogonal. Similarly,
off-diagonal geometric phases \cite{manini00} are undefined for cyclic
evolution. Thus, it appears to be a general fact that geometric phases
have nodal points at which they are undefined. Nevertheless, these
phases probe the geometry of the ray space, which is free from any
singularities. This observation raises the question if the concept of
geometric phase can be modified so that it reflects the absence of
singularities in ray space. Here, we propose such a modified concept,
which we shall call nodal free geometric phases.

Recently, geometric phases have been suggested to play a role in the
design of quantum gates that may be robust to certain kinds of
errors. This idea was first put forward in the case of cyclic
adiabatic evolution \cite{zanardi99}, and was subsequently 
implemented in the nonadiabatic \cite{xiang02} and noncyclic 
\cite{friedenauer03,kult06} contexts. The desired geometric 
properties may also occur when the dynamical phase becomes 
proportional to the geometric phase, leading to the notion of
unconventional geometric quantum computation \cite{zhu03}. Here, we
propose the nodal free geometric phases as another conceptual basis
for the construction of quantum gates.

Let $U(s)$, $s\in [0,1]$, be an arbitrary one-parameter family of
unitary operators acting on an $N$-dimensional Hilbert space
$\mathcal{H}$. We require that $U(0)=\hat{1}$, i.e., the identity 
operator on $\mathcal{H}$. This family is said to be parallel 
transporting if there exists an orthonormal basis 
$\psi =\{ \ket{\psi_k} \}_k$ of $\mathcal{H}$ such that each 
$U(s) \ket{\psi_k}$ satisfies the standard parallel transport 
condition \cite{anandan87,anandan88a,anandan88b} 
\begin{eqnarray} 
\bra{\psi_k} U^{\dagger} (s) \dot{U} (s) \ket{\psi_k} = 0.
\end{eqnarray}
Not all $U(s)$ has this property, e.g., for a qubit $(N=2)$, 
$e^{-is\sigma_z}$ is parallel transporting (any pair of 
orthogonal vectors that represent states on the equator of the 
Bloch sphere will do), while $e^{-is \sigma_x} e^{-is \sigma_z}$ 
is not.
 
Now, let $U_{\psi}^{\parallel} (s)$, $s\in [0,1]$ and 
$U_{\psi}^{\parallel} (0)= \hat{1}$, be a family of
unitary operators that parallel transport the orthonormal basis 
$\psi = \{ \ket{\psi_k}\}_k$. We are interested in the properties 
of $U_{\psi}^{\parallel} (1)$. Consider the $N \times N$ matrix
\begin{eqnarray} 
\boldsymbol{U}_{\psi}^{\parallel} = 
\begin{pmatrix}
\sigma_{11}  & \ldots & 
\sigma_{1N} \\
\vdots & \ddots & \vdots \\
\sigma_{N1} & \ldots & 
\sigma_{NN} 
\end{pmatrix} , 
\end{eqnarray}
where $\sigma_{kl} = \bra{\psi_k} U_{\psi}^{\parallel} (1)
\ket{\psi_l}$, i.e., $\boldsymbol{U}_{\psi}^{\parallel}$ is the matrix
representation of $U_{\psi}^{\parallel} (1)$ with respect to
$\psi$. Since $\boldsymbol{U}_{\psi}^{\parallel}$ is a unitary matrix,
it has $N$ unit modulus eigenvalues. These eigenvalues coincide with
the standard geometric phase factors in cyclic evolution
\cite{aharonov87,anandan88c,wu94} if
$\boldsymbol{U}_{\psi}^{\parallel}$ is diagonal, i.e., when all
$U_{\psi}^{\parallel} (s) \ket{\psi_k}$ undergo cyclic evolution. In
the noncyclic case, $\boldsymbol{U}_{\psi}^{\parallel}$ has nonzero
off-diagonal elements. The corresponding eigenvalues are still gauge
invariant phase factors, i.e., invariant under transformations of the
form $ \ket{\psi_k} \rightarrow e^{i\alpha_k} \ket{\psi_k}$, but
differ from the standard geometric phase factors in noncyclic
evolution \cite{samuel88}. In particular, the eigenvalues are always
well-defined, i.e., there are no nodal points where they become
undefined. These eigenvalues are the nodal free geometric phase
factors of the set of paths $\{ \Pi [U_{\psi}^{\parallel} (s)
\ket{\psi_k}] \}_k$ in ray space, $\Pi$ being the projection map 
\cite{aharonov87}.

To prove the gauge invariance of the nodal free geometric phase
factors, we note that $\sigma_{kl} \rightarrow \sigma_{kl} e^{-i
(\alpha_k - \alpha_l ) }$, under $\ket{\psi_k} \rightarrow
e^{i\alpha_k} \ket{\psi_k}$. Thus, $\boldsymbol{U}_{\psi}^{\parallel}
\rightarrow \boldsymbol{V} \boldsymbol{U}_{\psi}^{\parallel}
\boldsymbol{V}^\dagger$, where $\boldsymbol{V}=\mbox{diag}[e^{-i
\alpha_1}, e^{-i \alpha_2}, \ldots , e^{-i \alpha_N}]$, from which
it follows that the eigenvalues are preserved. 

Similarly as for the standard \cite{samuel88} and off-diagonal 
\cite{manini00} geometric phase factors 
the nodal free geometric phase factors can be defined in terms of
gauge-invariant quantities.  The eigenvalues $\lambda$ of
$\boldsymbol{U}_{\psi}^{\parallel}$ are solutions of the secular
equation $\det (\boldsymbol{U}_{\psi}^{\parallel} - \lambda
\boldsymbol{1}) =0$, $\boldsymbol{1}$ being the $N \times N$ unit
matrix. Expanding the determinant yields an equation that involves
$\gamma_{j}^{(1)} =
\sigma_{jj}$ and $\gamma_{j_1 \ldots j_l}^{(l)}
\equiv \sigma_{j_1j_l} \cdots \sigma_{j_2j_1}$, $l=2,3\ldots,N$, 
which are the gauge invariant quantities that define the standard  
and off-diagonal geometric phase factors $\Phi
[\gamma^{(l)}]$, where $\Phi [z] = z/|z|$ for any nonzero
complex number $z$. For instance, for $N=3$ we have
\begin{eqnarray} 
 & & -\lambda^3 + \big( \gamma_1^{(1)} + \gamma_2^{(1)} + 
\gamma_3^{(1)} \big) \lambda^2 
\nonumber \\ 
 & & - \big( \gamma_1^{(1)} \gamma_2^{(1)} + \gamma_2^{(1)} 
\gamma_3^{(1)} + \gamma_3^{(1)} \gamma_1^{(1)} -  \gamma_{12}^{(2)} 
- \gamma_{23}^{(2)} - \gamma_{31}^{(2)} \big) \lambda 
\nonumber \\ 
 & & + 
\gamma_{123}^{(3)} + \gamma_{132}^{(3)} - 
\gamma_{12}^{(2)} \gamma_3^{(1)} - \gamma_{23}^{(2)} \gamma_1^{(1)} - 
\gamma_{31}^{(2)} \gamma_2^{(1)} 
\nonumber \\ 
 & & + \gamma_1^{(1)} \gamma_2^{(1)} 
\gamma_3^{(1)} = 0.  
\end{eqnarray}
The solutions of this equation are fully determined by all the 
$\gamma$'s. 

One may derive some results for the off-diagonal geometric phases
using the above nodal free scenario. We first note that a central
motivation for Manini and Pistolesi \cite{manini00} to introduce the
concept of off-diagonal geometric phases was to retain geometric phase
information of the evolution in cases where the standard geometric
phase factors were undefined. In the present context, we can indeed
see that this must always be the case since if, for a given Hilbert
space dimension $N$, all $\gamma$'s happened to vanish and thereby all
$\Phi[\gamma]$ were undefined, the secular equation would reduce to
$\lambda^N = 0$, which contradicts the fact that $|\lambda|=1$ (see
also Ref. \cite{kult07}). Furthermore, one may see that the
$\lambda^0$ coefficient of the eigenvalue equation has the structure
$\gamma^{(N)} + \gamma^{(N-1)} \gamma^{(1)} + \gamma^{(N-2)} \big[
\gamma^{(1)} \big]^2 + \gamma^{(N-2)} \gamma^{(2)} + \ldots + \big[
\gamma^{(1)} \big]^N$. This term must be nonzero, since if it would
vanish, then $\lambda =0$ would be a solution of the eigenvalue
equation. It follows, in particular, that $\gamma^{(l\neq n)}=0$ for 
some fixed $n$ is only possible for the `extremal'
$n=1$ and $n=N$.

Next, we consider the relation between the nodal free phase factors
and the cyclic geometric phase factors \cite{aharonov87,anandan88c,wu94}. 
Let $\ket{\phi_k}$ be orthonormal eigenvectors of $U_{\psi}^{\parallel}
(1)$. These are cyclic states, i.e., $U_{\psi}^{\parallel} (1)
\ket{\phi_k} = \lambda_k \ket{\phi_k}$, where $\lambda_k$ are the 
nodal free geometric phase factors. Note that the eigenvectors are 
in general not parallel transported by $U_{\psi}^{\parallel} (s)$. 
The nodal free phase factors $\lambda_k$ differ from the standard 
cyclic geometric phase factors $e^{i\beta_k}$
\cite{aharonov87,anandan88c,wu94} associated with the paths 
$\Pi [U_{\psi}^{\parallel} (s) \ket{\phi_k}]$ in ray space. To see
this we first note that given the family of unitaries
$U_{\psi}^{\parallel} (s)$ the Schr\"odinger equation uniquely
determines the family of Hamiltonians $H(s) = i 
\dot{U}_{\psi}^{\parallel} (s){U_{\psi}^{\parallel}}^{\dagger}(s)$ 
that generates $U_{\psi}^{\parallel} (s)$.  If we let $\tau_k$ be the 
phase associated with the nodal free phase factor, i.e., 
$\lambda_{k} = e^{i\tau_k}$, we can use the technique in 
Ref. \cite{aharonov87} (see Eq. (3) in Ref. \cite{aharonov87})
to find that the cyclic geometric phases $\beta_{k}$ of
$|\phi_{k}\rangle$ are
\begin{equation}
\beta_{k} = \tau_k + 
\int_{0}^{1}\langle \phi_{k}(s)|H(s)|\phi_{k}(s)\rangle ds,
\end{equation}
which we can rewrite as
\begin{equation} 
e^{i \beta_k} =  
\lambda_{k}e^{i\int_0^1 \bra{\phi_k} 
H(s) \ket{\phi_k} ds} .  
\end{equation}
This establishes an explicit relation between the nodal free 
geometric phase factors and the geometric phase factors associated 
with the states that undergo cyclic evolution for the family 
$U_{\psi}^{\parallel} (s)$. As can be seen, these two differ by the 
dynamical phase factors of the eigenvectors $\ket{\phi_k}$. It 
is to be noted that, since the family of Hamiltonians is uniquely 
determined, these dynamical phase factors do not introduce any 
ambiguity in the definition of the nodal free geometric phases. 

As a final note on the general properties of the nodal free geometric
phases, we point out that these can be measured in interferometry. A
beam of particles with some internal degree of freedom (e.g., spin or
polarization) prepared in the internal state $\ket{\varphi}$ is
splitted by a 50-50 beam-splitter into two beams. In one of the
resulting beams, the internal state is transformed by the unitary
operators $U_{\psi}^{\parallel} (s), \ s\in [0,1]$, that parallel
transport the basis $\psi = \{ \ket{\psi_k} \}_k$, and a variable U(1)
phase shift $e^{i\chi}$ is applied to the other beam. The two beams
are brought back to interfere at a second beam-splitter. The resulting
interference pattern is determined by the complex-valued quantity
$F(\varphi) = \bra{\varphi} U_{\psi}^{\parallel} (1) \ket{\varphi}$ in
that the intensity measured in one of the output beams reads
\cite{wagh95}
\begin{eqnarray}
\mathcal{I} & \propto & 
1 + \big| F(\varphi) \big| \cos \big[ \chi - \arg F(\varphi) \big] . 
\end{eqnarray}
Explicitly, 
\begin{eqnarray} 
F(\varphi) = 
\sum_k | \bra{\phi_k} \varphi \rangle|^2 \lambda_k ,   
\end{eqnarray}
where $\sum_k \big| \bra{\phi_k} \varphi \rangle \big|^2 = 1$,
i.e., the interference function $F(\varphi)$ is a convex combination
of the nodal free geometric phase factors. It follows that the
interference oscillations are shifted by the nodal free geometric
phases $\arg \lambda_k$ if and only if $\ket{\varphi}$ coincide with
$\ket{\phi_k}$, in case of which $\big| F(\varphi) \big| = 1$. Thus,
the nodal free geometric phases can be obtained by varying the input
internal state $\varphi$ until unit visibility is attained.

To illustrate the theory with a specific example, let us analyze 
the nodal free geometric phases in the qubit case in some detail. 
Let $\eta = \big| \bra{\psi_1} U_{\psi}^{\parallel} (1) \ket{\psi_1}
\big| = \big| \bra{\psi_2}  U_{\psi}^{\parallel} (1) \ket{\psi_2} \big|$ 
measure the degree of cyclicity and let $\Omega$ be the solid angle
enclosed by the path traced out by the Bloch vector and the shortest
geodesics connecting the end-points. In terms of these quantities, we
have $\gamma_1^{(1)} = (\gamma_2^{(1)})^{\ast} = \eta e^{-i\Omega /2}$
and $\gamma_{12}^{(2)} = \eta^2-1$. Thus, if $\eta = 0$ then the
geometric phase factors $\Phi [\gamma_1^{(1)}]$ and $\Phi
[\gamma_2^{(1)}]$ are undefined, but the off-diagonal phase factor
$\Phi [\gamma_{12}^{(2)}] = -1$.  If $\eta = 1$ then the off-diagonal
geometric phase factor $\Phi [\gamma_{12}^{(2)}]$ is undefined and
$\Phi [\gamma_1^{(1)}] = \big( \Phi [\gamma_2^{(1)}]\big)^{\ast} = 
e^{-i\Omega /2}$. On the other hand, the nodal free geometric phase 
factors,  i.e., the eigenvalues of $\boldsymbol{U}_{\psi}^{\parallel}$, 
are always defined. The secular equation reads $\lambda^{2} - 
(\gamma_{1}^{(1)} + \gamma_{2}^{(2)}) \lambda +\gamma_{1}^{(1)} 
\gamma_{2}^{(1)} - \gamma_{12}^{(2)} =0$, which gives the 
eigenvalues
\begin{equation}
\lambda_{\pm}  = \eta\cos(\Omega/2) \pm 
i\sqrt{1-\eta^{2}\cos^{2}(\Omega/2)}  
\label{eq:qubitev}  
\end{equation}
with corresponding eigenstates $\ket{\phi_{\pm}}$.  
We note that $1\geq \eta^{2}\cos^{2}(\Omega/2)$ and
$|\lambda_{\pm}|=1$. For `bit flip' evolution ($\eta = 0$), the
geodesically closed solid angle $\Omega$ is undefined since there is
an infinite number of ways to close the path by a geodesics, and all
these closures yield different solid angles.  However, whatever
geodesic closure we choose, the nodal free geometric phase factors are
given by $\pm i$. For cyclic evolution ($\eta = 1$), we obtain the two
eigenvalues $\lambda_{\pm} = \cos(\Omega/2) \pm
i|\sin(\Omega/2)|$. Hence, as expected we find the two cyclic
geometric phase factors $e^{i\Omega/2}$ and $e^{-i\Omega/2}$. More
precisely, $\lambda_{+,-} = \Phi[\gamma_{2,1}^{(1)}]$ for
$\sin(\Omega/2)>0$ and $\lambda_{+,-} = \Phi[\gamma_{1,2}^{(1)}]$ for
$\sin(\Omega/2)<0$, i.e., the labeling of the eigenstates depends on
the sign of $\sin(\Omega/2)$ (note that $\lambda_+ = \lambda_-$ when
$\sin(\Omega/2) = 0$, which allows for the 'flip' of the labeling).

\begin{figure}
\includegraphics[width = 15cm]{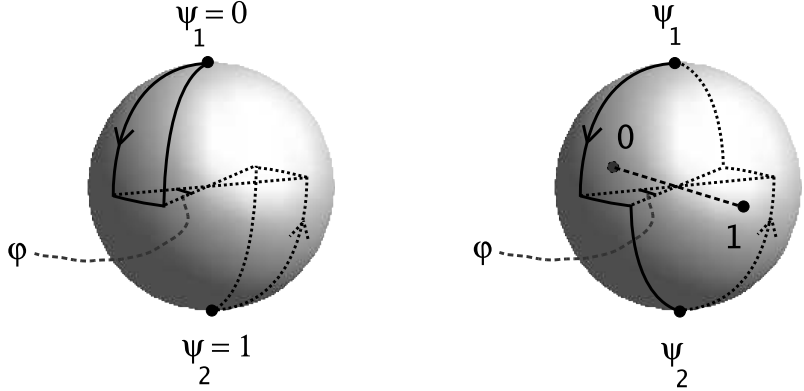}
\caption{\label{fig:fig1} Paths on Bloch sphere for phase shift gate
(left panel) and phase flip gate (right panel), based on the standard
and nodal free geometric phases, respectively. These paths can be
realized for instance by applying magnetic fields sequentially in
three different directions to a spin$-\frac{1}{2}$ particle.  The
phase shift gate is determined by the solid angle $\Omega =
\varphi$. Here, the computational and parallel transported bases
coincide. The phase flip gate depends only on the off-diagonal
geometric phase $\arg\gamma_{12}^{(2)}$. This phase equals half the
solid angle enclosed by the loop sequentially traced out by the path
starting at $\ket{\psi_1}$, followed by the path starting at
$\ket{\psi_2}$, i.e., $\arg \gamma_{12}^{(2)} =\pi$
\cite{manini00}. Thus, the nodal free geometric phase factors $\pm
\sqrt{\gamma_{12}^{(2)}}=\pm i$ that define this gate are $\varphi$
independent. Note, though, that the precise location of the
computational basis indicated by `$0$' and `$1$', on the equator of
the Bloch sphere, depends on $\varphi$. One can show that $\ket{0} =
\frac{1}{\sqrt{2}} \big( \ket{\psi_1} + e^{-i(\varphi - \pi)/2}
\ket{\psi_2} \big)$ and $\ket{1} = \frac{1}{\sqrt{2}} \big(
\ket{\psi_1} + e^{-i(\varphi + \pi)/2} \ket{\psi_2} \big)$ for this
phase flip gate.}
\end{figure}

The cyclic geometric phases arising from parallel transport have been
considered for implementations of phase gates \cite{li02}. In such
applications, the parallel transported and computational bases
coincide. The nodal free geometric phases provide an alternative
implementation of phase gates, if we let the computational basis
coincide with the eigenstates of $U^{\parallel}_{\psi}(1)$. Here, 
the computational basis in general does not coincide with the parallel
transported basis. In the single qubit case, considered above, we
choose the computational basis $\ket{0} \equiv
\ket{\phi_-}$ and $\ket{1}\equiv
\ket{\phi_+}$. The nodal free geometric phases may thus be
interpreted as a realization of the one-qubit phase gate $U =
\lambda_- \ket{0}\bra{0} + \lambda_+ \ket{1}\bra{1}$.  Such a phase
gate is fully determined by the quantities
$\gamma_{1}^{(1)},\gamma_{2}^{(1)},$ and $\gamma_{12}^{(2)}$ via the
eigenvalue equation. In the special case of cyclic evolution of 
the parallel transported basis, $U$ reduces to the standard nonadiabatic
geometric phase gate
$\gamma_1^{(1)}\ket{0}\bra{0}+\gamma_2^{(1)}\ket{1}\bra{1} =
e^{-i\Omega/2}\ket{0}\bra{0}+e^{i\Omega/2}\ket{1}\bra{1}$, which is
sensitive to changes in the solid angle $\Omega$.  On the other hand,
the off-diagonal geometric phase factor $\Phi[\gamma_{12}^{(2)}]$ is
either undefined (for cyclic evolution $\eta =1$) or $-1$ (for $0\leq
\eta < 1$). This $\Omega$ independence can be understood geometrically
by interpreting the second order off-diagonal geometric phase in terms
of a loop that always encloses the solid angle $2\pi$. This property
has been demonstrated experimentally for neutron spin
\cite{hasegawa01} and suggests that the off-diagonal geometric phases
may be a useful component for the design of quantum gates.  Since
$\gamma_{12}^{(2)} = \gamma_{21}^{(2)}$, it seems at first sight that
only a trivial phase gate can be implemented based on the off-diagonal
geometric phase. However, in the nodal free geometric phase scenario 
with $\eta = 0$, we obtain
\begin{eqnarray}
iZ = \sqrt{\gamma_{12}^{(2)}}\ket{0}\bra{0} - 
\sqrt{\gamma_{12}^{(2)}}\ket{1}\bra{1} = i\ket{0}\bra{0} - 
i \ket{1}\bra{1} ,     
\end{eqnarray} 
which, up to the unimportant overall phase factor $i$, is a nontrivial
phase flip gate. Thus, the nodal free geometric phase concept makes it
possible to use off-diagonal geometric phases in the design of quantum
gates. The robustness with respect to changes in $\Omega$ suggests
that this gate implementation would have an inherent stability against
noise in this parameter. Realizations of paths on the Bloch sphere
leading to the phase shift and phase flip gates using the standard and
nodal free geometric phases, respectively, are shown in
Fig. (\ref{fig:fig1}).

To develop this idea further, we consider two-qubit geometric 
phase gates, in case of which the relevant Hilbert space is 
four-dimensional. One may seek for a similar path independence 
as in the above phase flip gate by looking for 
implementations that only involve the fourth order ($l=4$) 
off-diagonal geometric phases. To this end, we assume that 
$U_{\psi}^{\parallel} (1)$ takes the matrix form 
\begin{eqnarray} 
\boldsymbol{U}_{\psi}^{\parallel} (1) = 
\begin{pmatrix}
0  & \sigma_{12} & 0 & 0 \\ 
0  & 0 & \sigma_{23} & 0 \\
0  & 0 & 0 & \sigma_{34} \\ 
\sigma_{41} & 0 & 0 & 0 
\end{pmatrix}  
\end{eqnarray}
in the parallel transported basis. The eigenvalue equation reads
$\lambda^4-\gamma_{1234}^{(4)}=0$, yielding the nodal free geometric
phase factors $\lambda_k = e^{ik\pi/2} \big( \gamma_{1234}^{(4)}
\big)^{1/4}$, $k=0,\ldots,3$. The parallel transporting nature of
the family $U_{\psi}^{\parallel} (s)$ implies that 
$U_{\psi}^{\parallel}(1)\in$ SU(4), i.e., 
$\det U_{\psi}^{\parallel}(1) =1$. Thus, $\gamma_{1234}^{(4)}=-1$
so that $\lambda_k=e^{i(2k+1)\pi/4}$, $k=0,\ldots ,3$. Let us assume
that there is a physically natural tensor product decomposition of the
Hilbert space such that the eigenvectors of $U_{\psi}^{\parallel}(1)$
coincide with the computational product basis. For instance, if
$\ket{\phi_0}=\ket{00},\ket{\phi_1}=\ket{01},\ket{\phi_2}=
\ket{11},\ket{\phi_3}=\ket{10}$, we obtain the conditional gate
\begin{eqnarray}
B = e^{i\frac{\pi}{4}} \ket{00} \bra{00} + 
e^{i\frac{3\pi}{4}} \ket{01} \bra{01} + 
e^{i\frac{5\pi}{4}} \ket{11} \bra{11} +  
e^{i\frac{7\pi}{4}} \ket{10} \bra{10} .  
\end{eqnarray}
On the other hand, if $\ket{\phi_0}=\ket{00},\ket{\phi_1}=\ket{11},
\ket{\phi_2}=\ket{10},\ket{\phi_3}=\ket{01}$, we obtain the 
product gate 
\begin{eqnarray}
Z \otimes S = \big( \ket{0} \bra{0} - 
\ket{1} \bra{1} \big) \otimes \big( e^{i \pi/4}\ket{0} \bra{0} - 
e^{-i\pi/4}\ket{1} \bra{1} \big) , 
\end{eqnarray}
where $S$ is the phase or $\pi /4$ gate. We may continue to apply this
idea for more qubits. In particular, we notice that the procedure
admits a realization of the $\pi/8$ gate using three qubits, based
entirely on the extremal $\gamma_{12\ldots 8}^{(8)}$. This $\pi/8$ gate
is an important ingredient to achieve universal fault tolerant quantum
computation \cite{boykin00}.
  
In conclusion, we introduce a family of geometric phases for parallel
transporting unitary evolutions. These are explicitly constructed from
gauge invariant quantities, but nevertheless do not have the nodal
structure concomitant to the noncyclic and off-diagonal geometric 
phases. We furthermore show that the nodal free geometric phases 
have an operational meaning in the sense that they can be measured 
using interferometry. We point out that nodal free geometric phases 
could be useful in geometric quantum computation as they may show 
robustness features that are not shared by the standard geometric 
phases. 
\vskip 0.5 cm  
M.E. acknowledges financial support from the Leverhulme Trust. 
E.S. acknowledges financial support from the Swedish Research Council. 
J.{\AA}. wishes to thank the Swedish Research Council for financial 
support and the Centre for Quantum Computation at DAMTP, Cambridge, 
for hospitality. The work by M.E. and J.{\AA}. was supported by the 
European Union through the Integrated Project QAP (IST-3-015848), 
SCALA (CT-015714), SECOQC and the QIP IRC (GR/S821176/01).


\begin{thebibliography}{99}  
\bibitem{samuel88} J. Samuel, R. Bhandari, 
Phys. Rev. Lett. 60 (1988) 2339.
\bibitem{manini00} N. Manini, F. Pistolesi,
Phys. Rev. Lett. 85 (2000) 3067.
\bibitem{zanardi99} P. Zanardi, M. Rasetti,
Phys. Lett. A 264 (1999) 94.
\bibitem{xiang02} W. Xiang-Bin, M. Keiji
Phys. Rev. Lett. 87 (2002) 097902.
\bibitem{friedenauer03} A. Friedenauer, E. Sj\"oqvist,  
Phys. Rev. A 67 (2003) 024303.
\bibitem{kult06} D. Kult, J. {\AA}berg, E. Sj\"oqvist, 
Phys. Rev. A 74 (2006) 022106.
\bibitem{zhu03} S.-L. Zhu, Z.D. Wang,
Phys. Rev. Lett. 91 (2003) 187902.
\bibitem{anandan87} J. Anandan, L. Stodolsky,  
Phys. Rev. D 35 (1987) 2597.
\bibitem{anandan88a} J. Anandan,  
Phys. Lett. A 129 (1988) 201.
\bibitem{anandan88b} J. Anandan, Y. Aharonov,  
Phys. Rev. D 38 (1988) 1863.
\bibitem{aharonov87} Y. Aharonov, J. Anandan,  
Phys. Rev. Lett. 58 (1987) 1593.
\bibitem{anandan88c} J. Anandan,  
Phys. Lett. A 133 (1988) 171.
\bibitem{wu94} L.-A. Wu, 
Phys. Rev. A 50 (1994) 5317. 
\bibitem{kult07} D. Kult, J. {\AA}berg, E. Sj\"oqvist, 
Europhys. Lett. 78 (2007) 60004. 
\bibitem{wagh95} A.G. Wagh, V.C. Rakhecha, 
Phys. Lett. A 197 (1995) 107. 
\bibitem{li02} X.-Q. Li, L.-X. Cen, G. Huang, L. Ma, Y. Yan, 
Phys. Rev. A 66 (2002) 042320.  
\bibitem{hasegawa01} Y. Hasegawa, R. Loidl, M. Baron, G. Badurek, 
H. Rauch, 
Phys. Rev. Lett. 87 (2001) 070401. 
\bibitem{boykin00} P.O. Boykin, T. Mor, M. Pulver, V. Roychowdhury, 
F. Vatan, 
Inform. Process. Lett. 75 (2000) 101.  
\end{thebibliography}
\end{document}